\documentclass[apj]{emulateapj}

\usepackage{multirow}
\usepackage{booktabs}
\usepackage{array}

\DeclareMathAlphabet{\mathsc}{OT1}{cmr}{m}{sc}
\def\testbx{bx}
\DeclareRobustCommand{\ion}[2]{
\relax\ifmmode
\ifx\testbx\f@series
{\mathbf{#1\,\mathsc{#2}}}\else
{\mathrm{#1\,\mathsc{#2}}}\fi
\else\textup{#1\,{\mdseries\textsc{#2}}}
\fi}
\newcommand{\synow} {\textsc{synow}}
\newcommand{\epm} {\textsc{epm}}
\newcommand{\ned} {\textsc{ned}}
\newcommand{\scm} {\textsc{scm}}
\newcommand{\nlte} {\textsc{ nlte}}
\newcommand{\lte} {\textsc{lte}}
\newcommand{\cmfgen} {\textsc{cmfgen}}
\newcommand{\seam} {\textsc{seam}}
\newcommand{\iraf} {\textsc{iraf}}
\newcommand{\splot} {\textsc{splot}}

\newcommand{\ve}[2]{$#1 \pm #2$}
\newcommand{\figcap}[2]{Same as Fig. \ref{fig:epm_99gi}, but for SN #1.}
\newcommand{\figcapFix}[2]{Same as Fig. \ref{fig:epm_99gi}, but for SN #1 with fixed explosion epoch at JD #2.}

\newcommand{\Feii} {\ion{Fe}{ii}}

\newcommand{\Baii} {\ion{Ba}{ii}}
\newcommand{\Nai} {\ion{Na}{i}}

\newcommand{\Scii} {\ion{Sc}{ii}}

\newcommand{\Tiii} {\ion{Ti}{ii}}
\newcommand{\Hei} {\ion{He}{i}}

\newcommand{\ebv}{\mbox{$E(B-V)$}}

\newcommand{\msun}{\mbox{$M_{\odot}$}}

\newcommand{\kms}{\mbox{$\rm{\,km\,s^{-1}}$}}

\newcommand{\cobalt}{\mbox{$^{56}$Co}}

\shorttitle{EPM distances to eight type IIP supernovae}
\shortauthors{BOSE \& KUMAR}

\begin{document}

\title{Distance determination to eight galaxies using expanding photosphere method}

\author{
Subhash Bose$^{\star}$\altaffilmark{} and
Brijesh Kumar\altaffilmark{}
}

\email{$^\star$email@subhashbose.com, bose@aries.res.in}

\altaffiltext{}{Aryabhatta Research Institute of Observational Sciences, Manora
    Peak, Nainital 263 002, India}

\begin{abstract}
Type IIP supernovae are recognized as independent extragalactic distance indicators, however,
keeping in view of the diverse nature of their observed properties as well as the availability of
good quality data, more and newer events need to be tested for their applicability as a reliable
distance indicators. We use early photometric and spectroscopic data of eight type-IIP SNe to
derive distances to their host galaxies using the expanding photosphere method (\epm).
For five of these, \epm\, is applied for the first time. In this work, we improved \epm\, application
by using \synow\, estimated velocities and by semi-deconvolving the broadband filter responses while
deriving color temperatures and black-body angular radii.  We find that the derived \epm\, distances
are consistent with that derived using other redshift independent methods.

\end{abstract}

\keywords{supernovae: general $-$ \epm}

\section{Introduction} \label{int}

The hydrostatic nuclear burning phases of massive stars with initial masses
greater than about 8 \msun\ results in a onion-skin like stratification of
nucleosynthesis yields as well as the unprocessed material consisting of iron core and
successive zones of lighter elements up to helium and
hydrogen \citep{1996snih.book.....A,2011RPPh...74i6901J}.
It is understood that supernovae (SNe) explosions mark end stages in the life of
these stars \citep{2003ApJ...591..288H,2009ARA&A..47...63S}, and the explosion results in
collapse of iron core into a stellar mass compact object followed by shock-driven heating
and expulsion of outer stellar envelope, although, the exact
mechanism of explosion and the chemical yields from explosive nucleosynthesis are not fully
understood \citep{1995ApJS..101..181W,2012ARNPS..62..407J,2013RvMP...85..245B}.

A majority of core-collapse events showing
hydrogen lines in their optical spectra are classified as type II SNe \citep{1997ARA&A..35..309F},
and their progenitors are thought to have retained enough hydrogen until the time of explosion.
About ninety percent of all type II events are sub-classified as
IIP \citep{2009MNRAS.395.1409S, 2010ApJ...721..777A, 2011MNRAS.412.1522S}.
The V-band light curve of IIP SNe are described by a fast rise (up to 10-15 days post explosion);
a long plateau
phase for about hundred days which is sustained by cooling down of shock-heated expanding ejecta
by hydrogen recombination and then an exponential decline powered by radioactive decay of
newly formed \cobalt\, \citep{bose13aw}. Study of pre-supernova stars from the archival
pre-explosion images proves beyond doubt that the progenitors of IIP SNe are red
supergiant stars \citep{2009MNRAS.395.1409S,2013arXiv1304.4967P}.

Observations of type IIP SNe have been used to determine distances to their host galaxies using
expanding photosphere method (\epm), which is a variant of Baade-Wesselink method, developed and
implemented
first by \citet{kirshner74} for two SNe. The \epm\ provides an estimate of cosmological
distances, independent of extragalactic distance ladder, and offers alternative to verify results
obtained using other tools e.g. SN Ia. \cite{1992ApJ...395..366S,1994ApJ...432...42S}
applied \epm\ to several IIP SNe out to 180 Mpc to constrain value of Hubble constant (H$_{\rm 0}$).
\cite{e96} quantified the dilution factors of supernova atmospheres relative
to black-body function and gave a firm theoretical foundation to \epm. However, there have
been discrepancies in the distances derived using \epm, e.g. a value in the range 7 to 8 Mpc
is obtained for SN 1999em \citep{h01, 2002PASP..114...35L, 2003MNRAS.338..939E}, while a
value of $11.7\pm1.0$ Mpc is derived using Cepheids \citep{2003ApJ...594..247L}.
Subsequently, spectral-fitting expanding atmosphere method (\seam) employing the
full \nlte\ supernova model atmosphere codes have been used to derive
distances to SN 1999em \citep{2004ApJ...616L..91B,2006A&A...447..691D} and the
estimated value was found to be in fair agreement with the Cepheid distance. However, the \seam\
is computationally intensive and it can only be applied to events having high signal-to-noise
spectra at early phases. The \epm\ need to be explored further.

\cite{2009ApJ...696.1176J}
derived \epm\ distances to 12 IIP SNe using two sets of SN atmospheres, three filter
subsets, the photospheric velocity estimated from Doppler-shifts of spectral lines
and they found variation in \epm\ distances up to 50\% depending on the models
and subsets used. Recently, \cite{2012A&A...540A..93V} applied \epm\ to SNe 2005cs and 2011dh,
both in M51 and both having densely sampled light-curves and spectra and they derived distance
in good agreement with that in NED database. \cite{takats12} applied \epm\ to 5 IIP SNe
and found that photospheric velocity estimated using \synow\ models of spectral lines are preferred.

Due to their high intrinsic luminosity, type IIP SNe have been detected out to z=0.6 and are
expected to be more abundant at higher redshifts \citep{2006ApJ...651..142H}.
After finding a correlation between plateau luminosity and the photospheric velocity,
\cite{hamuy02} first established IIP SNe as standardizable candles. This standard candle method (\scm)
is consistent with red supergiants as their progenitors. Using model light curves of
IIP SNe, \cite{2009ApJ...703.2205K} gave a firm
theoretical basis to the tight relationship between luminosity and expansion velocity, though they
found a sensitivity to progenitor metallicity and mass. \cite{2010ApJ...715..833O} applied \scm\
to 37 nearby (z $<$ 0.06) SNe with relative distance precision of 12-14\%, though they
found systematic differences between distances derived using \epm\ and \scm.

The observed mid-plateau properties of type II-P SNe form a sequence from
subluminous $M_{\rm V} \sim -15$ mag, low-velocity $v \sim$ 2000 \kms\ to
bright $\sim -18$ mag, high-velocity $\sim$ 8000 \kms\ events \citep{2003ApJ...582..905H}.
Recently, a spectroscopically subluminous IIP showing light curve properties similar
to a normal luminosity event have also been observed, e.g. SN 2008in \citep{roy11}
and SN 2009js \citep{2013ApJ...767..166G}. Several bright events showing signs of
circumstellar interaction have been observed, e.g. SNe 2007od \citep{inserraOD11}
and 2009bw \citep{inserra12}. The main factors governing the observed properties are nature
and environments of progenitors. In view of diversity in the properties of IIP SNe,
as well as the availability of good quality data for several events in the literature,
more and newer events need to be tested for its applicability as reliable distance indicators.
In this work, we  extend the \epm\ analysis to eight type IIP SNe
having sufficient early time
photometric and spectroscopic data to test the full applicability of \epm\
and know the limitations and strength of the method.

The paper is organized as follows. The basic ingredients of the \epm\ are briefly described
in \S\ref{sec:met}. The sample and data are given in \S\ref{sec:dat}. The \epm\ analysis,
sources of errors and results are presented and discussed in \S\ref{sec:epm},
followed by discussions on individual events and summary in \S\ref{sec:dis}.

\section{method} \label{sec:met}

The expanding photosphere method (EPM) is fundamentally a geometrical technique \citep{kirshner74,1992ApJ...395..366S}, in which we
compare the linear radii determined from the velocity of supernova expansion with that of angular
radii estimated by fitting blackbody to the observed supernova fluxes at different epochs.
For extragalactic supernovae, it is not possible to measure radii directly as they are seen
as point sources, however, we may relate linear radius $R$ and angular radius $\theta$  as
$\theta=R/D$, where $D$ is distance to the supernova.
Furthermore, assuming a spherically symmetric expansion of the photosphere moving with
velocity $v_{\rm ph}$  at time $t$ and neglecting other deceleration factor such as
gravity and interstellar medium, we may write the above geometric relation as
\begin{eqnarray}
t=D\left( \frac{\theta}{v_{\rm ph}} \right)+t_0
\label{eq:epm}
\end{eqnarray}
where $t_0$ is the time of explosion. We use this linear equation to determine $D$ and $t_0$.
Given $t_0$, we can estimate $D$ for each value of $\theta/v_{\rm ph}$ and alternatively,
the relation can also be solved to estimate unique values of $D$ and $t_0$. We note that
for many supernovae, the later is not known with sufficient precision and the method can also
be used to get an independent estimate of $t_0$ as well as to test the consistency of the
fitted parameters.

Thus, to derive distance by \epm, all we need are values of $v_{\rm ph}$
and $\theta$ at different $t$. The former is derived from low-resolution optical spectroscopic
data while the later is estimated from broad-band photometric data.

\subsection{Determination of $v_{\rm ph}$} \label{sec:vel_det}

The determination of expansion velocity of supernova at the photosphere $v_{\rm ph}$ at time $t$ is
a non-trivial issue and several approaches have been evolved in the
literature, see \citet{takats12} for a summary on merits and demerits of various approaches.
The photosphere represents the optically thick and ionized part of the
ejecta which emits most of the continuum radiation as a diluted blackbody and it is understood
to be located in a thin spherical shell where electron-scattering optical depth of photons is 2/3
\citep{dh05}. In type IIP SNe, no single measurable spectral feature is directly connected with the true
velocity of photosphere, however, during the plateau-phase, it is best represented by
blue-shifted absorption components of P-Cygni profiles of \Feii\ at 4924\AA, 5018\AA\ and 5169\AA.
In early-phase ($t\le$ 10-15 d) of SNe, the \Feii\ lines are either weak or absent and in such cases, the
\Hei\ 5876\AA\, line can be used to estimate photospheric velocity with an accuracy of 2--4\%
\citep[e.g.][]{vinko07,takats06}, however at later phases ($t>$ 20 d), \Hei\ lines disappear and \Nai~D lines
start to dominate in same spectral region. We can estimate velocities either by measuring Doppler-shift of
the absorption minima using \splot\ task of \iraf\ (denoted as $v_{\rm pha}$) or by modelling the
observed spectra with \synow\,($v_{\rm phs}$). We use both the methods in this work.

\synow\ \citep{1997ApJ...481L..89F,1999MNRAS.304...67F,branch02} is a highly parameterized
spectrum synthesis code with number of simplified assumptions: homologous expansion, spherical symmetry,
line formation is purely due to resonant scattering in which the radiative transfer equations are solved
by Sobolev approximation and the most important assumption is \lte\ atmosphere with a sharp photosphere
radiating like a blackbody. However, despite such simplified assumptions, the basic physics of
expanding photosphere is preserved which gives rise to P-Cygni profiles for each spectral line. As a
result, the underlying continuum of the synthetic spectra shall not match with observed ones because
of the obvious fact that the physics of the continuum is significantly different and definitely
not \lte\, but,
the P-Cygni profiles shall be well reproduced in synthetic spectra which is directly related to the
velocity of line formation layers. The \synow\ also has the potential to reproduce line blending
features in synthetic spectra, as in case of \Feii\ line, these are moderately contaminated by other
ions, among which most prominent ones are \Baii, \Scii\ and \Tiii.
\cite{takats12} have compared the velocities determined from \synow\ and \cmfgen\ as the later model
solves the \nlte\ radiation-transfer equations for expanding photosphere, and it has been shown that the
velocities from each of these model are very much consistent with each other.

\begin{table*}
  \centering
  \caption{Basic properties for supernovae and their host galaxies.}
   \begin{tabular}{lcccccl}
    \hline \hline
     ID      &     Host &  $v_{\rm rec}$ &         $t_{\rm ref}$& $E(B-V)_{\rm tot}$& $M_{\rm V}$      &   References         \\
      (SN)   &  galaxy  & (\kms)         &           (JD) &              (mag)&   (mag)          &                      \\
        (1)  &      (2) &             (3)&             (4)&               (5) &    (6)           &    (7)               \\
    \hline
    SN 1999gi&  NGC 3184&           552  &  1518.2$\pm$3.1&   0.21$\pm$0.09   &  -16.3           &  \cite{leon02}    \\
    SN 2004et&  NGC 6946&            45  &  3270.5$\pm$0.9&   0.41            &  -17.1           &  \cite{sahu06,takats12}     \\
    SN 2005cs&  NGC 5194&           463  &  3549.0$\pm$1.0&   0.05$\pm$0.02   &  -15.1           &  \cite{pasto06,pasto09,2007ApJ...662.1148B}    \\
    SN 2006bp&  NGC 3953&           987  &  3834.5$\pm$2.0&   0.40            &  -17.1           &  \cite{immler07,quimby07}   \\
             &          &                &                &                   &                  &  \cite{dessart08}   \\
    SN 2008in&  NGC 4303&          1567  &  4825.6$\pm$2.0&   0.10$\pm$0.10   &  -15.7           &  \cite{roy11}      \\
    SN 2009bw&  UGC 2890&          1155  &  4916.5$\pm$3.0&   0.31$\pm$0.03   &  -16.8           &  \cite{inserra12}  \\
    SN 2009md&  NGC 3389&          1308  &  5162.0$\pm$8.0&   0.10$\pm$0.05   &  -14.9           &  \cite{fraser11}   \\
    SN 2012aw&  NGC 3351&           778  &  6002.6$\pm$0.8&   0.07$\pm$0.01   &  -16.7           &  \cite{bose13aw}   \\
    \hline
    \end{tabular}
    \\
    Notes : The columns are
           (1) identification of SN;
		   (2) identification of supernova host-galaxy;
		   (3) recession velocity of the galaxy used for doppler correction;
		   (4) the reference epoch in JD since 2450000.0, these are adopted explosion epoch from corresponding literature;
		   (5) the total reddening $E(B-V)_{\rm tot}$ towards the line-of-sight of SN;
		   (6) appromximate absolute visual magnitude at $\sim$50 day;
		   (7) references for $t_{ref}$, $E(B-V)_{\rm tot}$,  $M_{\rm V}$, and the photometric and spectroscopic data.

  \label{tab:par-sne}
\end{table*}

\subsection{Determination of $\theta$} \label{sec:the_det}

In order to determine $\theta$ at time $t$, we assume that the supernova is radiating isotropically as a
blackbody and hence accounting for the conservation of radiative energy we may write,
\begin{equation}
 f_\lambda^{obs} = \theta^2 \pi B_\lambda(T_c) 10^{-0.4A_\lambda} \label{eq:model}
\end{equation}
where $B_\lambda(T_c)$ is Planckian blackbody function at color temperature $T_c$, $A_\lambda$ is
the interstellar extinction and $f_\lambda^{obs}$ is the observed flux.

In practice, the value of $f_\lambda^{obs}$ from expanding photosphere of a supernova has significant
departure from a true blackbody emission, for the thermalization layer from
which the thermal photons are generated is significantly deeper
than photospheric layer defining the last scattering ($\tau=2/3$) surface.
As a result, while comparing blackbody flux with that of $f_\lambda^{obs}$, the value of $\theta$
corresponds to the thermalization layer, whereas the value of $v_{\rm ph}$ to
the photospheric layer and hence to take care of this discrepancy,
the ``dilution factor" $\xi$ is introduced \citep{1981ApJ...250L..65W} as
\begin{equation}
\xi=\frac{R_{therm}}{R_{ph}}
\end{equation}
and rewrite the equation~\ref{eq:model} as,
\begin{equation}
 f_\lambda^{obs} = \xi^2_\lambda \theta^2 \pi B_\lambda(T_c) 10^{-0.4A_\lambda} \label{eq:model2}
\end{equation}
Here, $\xi$ is termed as
distance correction factor as the distance derived without accounting flux dilution will
be overestimated by a factor of $1/\xi$. In principle, $\xi$ depends on many
physical properties including chemical composition and density profile of the ejecta. However,
\cite{e96} have shown that $\xi$ behaves more or less as one-dimensional function of $T_c$ only.
The computation of $\xi$ requires realistic SN atmosphere models and to be compared with blackbody model,
this requires high computing power and detailed physics of SN atmosphere, which is beyond the scope of
this paper. However, with the advent of faster and powerful computing, it is possible to execute
such model codes. Till date, two prescription for dilution factor is available, given independently
by \citet{e96} and by \citet{dh05}, hereafter D05. An improved estimate of $\xi(T_c)$ based on the
models of \citet{e96} was provided by \citet{h01}, hereafter H01.
In this paper we use prescriptions of both H01 and D05.

In principle, the value of  $f_\lambda^{obs}$ should be obtained from accurate spectrophotometry. However,
due to easy availability, it is derived from the photometric data taken using broad-band filters.
Consequently, the broadband filter response is inherently embedded within the quoted magnitudes.
In order to remove the effect of filter response in observed flux $f_\lambda^{obs}$ when compared with
blackbody model $\pi B(\lambda',T_c)$ flux, we convolve the response function for
each pass-band filter with the blackbody model to obtain the synthetic model flux.
If $\Re_\lambda(\lambda')$ be the normalized response function of a particular filter whose effective
wavelength is $\lambda$, then the convolved synthetic flux $b_\lambda$ is,
\begin{equation}
b_\lambda(T_c) = \int^\infty_0 \Re_\lambda(\lambda') \pi B(\lambda',T_c) d \lambda' \label{eq:fil_res}
\end{equation}
Hence the blackbody flux is replaced with convolved blackbody
flux $b_\lambda$ for each filter and equation~ \ref{eq:model} is rewritten as,
\begin{equation}
 f_\lambda^{\rm obs} = \xi^2_\lambda \theta^2 b_\lambda(T_c) 10^{-0.4A_\lambda} \label{eq:model_fin}
\end{equation}
In this paper we adopted the response function $\Re_\lambda$ for each of $UBVRI$ filters
from \cite{bessell90}.

In principle we should be able to use all filter passbands (UBVRI for optical) combination to apply
\epm. However, in practice all passbands are not suitable for such study; fast decaying magnitude
in U-band, makes the SN too faint for good observations, so, U band is generally opted out
from \epm; R-band is also unsuitable for \epm\ due to contamination from strong $H\alpha$ emission
in type II SNe. Hence only, three filter combinations are used for \epm\
study viz., \{BV\}, \{BVI\} and \{VI\} in combination to two set of dilution factors obtained from H01
and D05.

In reference to the preceding discussions, we are required to solve for $\theta$ and $T_c$. Hence
we construct $X$ using equation \ref{eq:model_fin} and recast in terms of broadband photometric
fluxes,
\begin{equation}
X = \sum_{j=BVI} [f_j^{obs} - \xi_j^2 \theta^2 b_j(T_c) 10^{-0.4A_j}]^2 \label{eq:chisq}
\end{equation}
On minimizing we obtain the quantities `$\theta\xi$' and `$T_c$' simultaneously, it is also to be
noted that $\xi(T_c)$ is itself the function of $T_c$. So we separate out $\theta$ by using the
known $\xi$ prescription for the particular filter combination used.

\begin{figure}
\centering
\includegraphics[width=8.5cm]{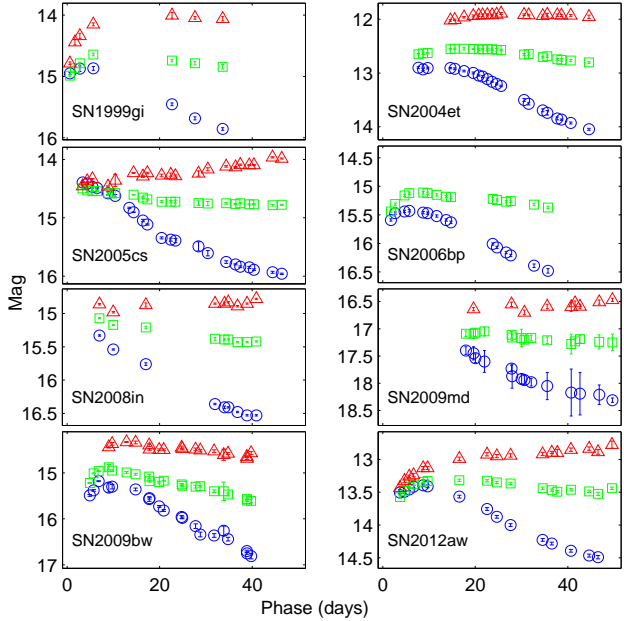}
\caption{The $BVI$ light curves of sample SNe. The colors blue, green and red
         indicate $B$, $V$ and $I$ bands respectively. The phases are in reference to the
         corresponding time of explosion $t_{\rm ref}$ adopted in table \ref{tab:par-sne}.}
\label{fig:photplot_all}
\end{figure}

\begin{table*}
  \caption{Photospheric velocities ($v_{\rm ph}$) of supernovae at different phases.}
  \label{tab:vel-sne}

  \begin{tabular}{c c c | c c c | c c c | c c c} \hline

\multicolumn{3}{c|}{SN 1999gi}    & \multicolumn{3}{c|}{SN 2004et}        &\multicolumn{3}{c|}{SN 2005cs}     &\multicolumn{3}{c}{SN 2006bp}     \\
Phase & $v_{\rm phs}$ &$v_{\rm pha}$& Phase & $v_{\rm phs}$     &$v_{\rm pha}$&Phase & $v_{\rm phs}$     &$v_{\rm pha}$     &Phase & $v_{\rm phs}$     &$v_{\rm pha}$\\ \hline
 4.7*  & 13.20 $\pm$  0.30  &12.79  &  11.1* &  8.90 $\pm$ 0.40  &  8.59  &   3.4*  &  6.30 $\pm$ 0.30  & 6.48  &   3.35 *&  13.70 $\pm$ 0.30 & 12.99 \\
 6.8*  & 10.30 $\pm$  0.40  &11.09  &  12.3* &  9.10 $\pm$ 0.40  &  9.54  &   4.4*  &  6.10 $\pm$ 0.20  & 6.04  &   6.30 *&  12.10 $\pm$ 0.20 & 11.33 \\
 7.8*  & 10.00 $\pm$  0.40  &11.07  &  13.0* &  9.40 $\pm$ 0.40  &  8.79  &   5.4*  &  5.70 $\pm$ 0.25  & 5.59  &   8.10 *&  11.50 $\pm$ 0.15 & 11.21 \\
 30.7  &  4.85 $\pm$  0.07  & 5.18  &  14.4* &  8.40 $\pm$ 0.40  &  9.04  &   8.4*  &  5.30 $\pm$ 0.30  & 5.16  &   10.10*&  10.55 $\pm$ 0.20 & 10.41 \\
 35.7  &  4.20 $\pm$  0.10  & 4.67  &  15.0* &  8.80 $\pm$ 0.20  &  8.31  &   8.8   &  5.30 $\pm$ 0.30  & 4.71  &   12.13 &   9.20 $\pm$ 0.40 & 10.01 \\
 38.7  &  4.05 $\pm$  0.10  & 4.47  &  16.0* &  8.00 $\pm$ 0.30  &  7.79  &   14.4  &  4.00 $\pm$ 0.30  & 3.76  &   16.10 &   9.00 $\pm$ 0.50 &  8.98 \\
 89.6  &  1.60 $\pm$  0.20  & 2.78  &  24.6  &  7.30 $\pm$ 0.40  &  6.41  &   14.4  &  4.10 $\pm$ 0.20  & 3.83  &   21.28 &   8.10 $\pm$ 0.10 &  7.69 \\
       &                    &       &  30.6  &  6.20 $\pm$ 0.20  &  5.69  &   17.4  &  3.60 $\pm$ 0.20  & 3.31  &   25.26 &   6.75 $\pm$ 0.30 &  6.33 \\
       &                    &       &  35.5  &  5.30 $\pm$ 0.30  &  4.98  &   18.4  &  3.60 $\pm$ 0.20  & 3.52  &   33.22 &   6.05 $\pm$ 0.20 &  5.63 \\
       &                    &       &  38.6  &  5.10 $\pm$ 0.15  &  4.98  &   22.5  &  3.20 $\pm$ 0.50  & 2.89  &   42.22 &   5.05 $\pm$ 0.10 &  4.79 \\
       &                    &       &  40.7  &  4.90 $\pm$ 0.25  &  4.86  &   34.4  &  2.40 $\pm$ 0.10  & 2.26  &   57.20 &   4.23 $\pm$ 0.05 &  4.78 \\
       &                    &       &  50.5  &  4.20 $\pm$ 0.25  &  4.28  &   36.4  &  2.25 $\pm$ 0.05  & 1.80  &         &                   &       \\
       &                    &       &  55.6  &  4.00 $\pm$ 0.25  &  3.85  &   44.4  &  1.95 $\pm$ 0.10  & 1.43  &         &                   &       \\
       &                    &       &  63.5  &  3.80 $\pm$ 0.10  &  3.66  &   61.4  &  1.40 $\pm$ 0.07  & 1.02  &         &                   &       \\
       &                    &       &        &                   &        &   62.4  &  1.35 $\pm$ 0.13  & 0.98  &         &                   &       \\ \hline \hline

\multicolumn{3}{c|}{SN 2008in}     & \multicolumn{3}{c|}{SN 2009bw}        &\multicolumn{3}{c|}{SN 2009md}      &\multicolumn{3}{c}{SN 2012aw}    \\
Phase & $v_{\rm phs}$     &$v_{\rm pha}$    & Phase & $v_{\rm phs}$     &$v_{\rm pha}$ &Phase & $v_{\rm phs}$     &$v_{\rm pha}$    &Phase & $v_{\rm phs}$     &$v_{\rm pha}$     \\ \hline
    7*& 6.10  $\pm$  0.10 & 5.72  & 4.0* & 8.90 $\pm$ 0.35  & 8.90  &   12  &   5.50 $\pm$ 0.40 & 6.22 &  7* & 11.20 $\pm$ 0.30  & 10.31   \\
   14 & 4.54  $\pm$  0.15 & 4.36  & 17.8 & 6.70 $\pm$ 0.40  & 6.84  &   15  &   5.30 $\pm$ 0.35 & 4.76 &  8* & 10.70 $\pm$ 0.30  &  9.55   \\
   54 & 2.80  $\pm$  0.07 & 2.66  & 19.8 & 6.15 $\pm$ 0.30  & 5.66  &   27  &   3.05 $\pm$ 0.10 & 2.92 &  12*&  9.00 $\pm$ 0.35  &  8.39   \\
   60 & 2.66  $\pm$  0.06 & 2.66  & 33.8 & 4.85 $\pm$ 0.20  & 4.68  &   48  &   2.05 $\pm$ 0.07 & 2.22 &  15*&  8.65 $\pm$ 0.30  &  8.14   \\
      &                   &       & 37.0 & 4.25 $\pm$ 0.25  & 4.37  &  100  &   0.85 $\pm$ 0.10 & 1.43 &  16 &  8.60 $\pm$ 0.25  &  8.29   \\
      &                   &       & 38.0 & 4.25 $\pm$ 0.15  & 4.56  &       &                   &      &  20 &  7.70 $\pm$ 0.20  &  7.46   \\
      &                   &       & 39.0 & 4.20 $\pm$ 0.10  & 4.25  &       &                   &      &  26 &  6.55 $\pm$ 0.20  &  6.25   \\
      &                   &       & 52.0 & 3.50 $\pm$ 0.20  & 3.60  &       &                   &      &  31 &  5.60 $\pm$ 0.10  &  5.51   \\
      &                   &       & 64.0 & 3.15 $\pm$ 0.10  & 3.16  &       &                   &      &  45 &  4.50 $\pm$ 0.06  &  4.47   \\
      &                   &       & 67.0 & 3.05 $\pm$ 0.10  & 3.08  &       &                   &      &  55 &  4.15 $\pm$ 0.08  &  4.02   \\
      &                   &       &      &                  &       &       &                   &      &  61 &  3.50 $\pm$ 0.05  &  3.68   \\
      &                   &       &      &                  &       &       &                   &      &  66 &  3.50 $\pm$ 0.10  &  3.61   \\ \hline

  \end{tabular}
  \newline
  Notes : Velocity derived using \synow\ is denotated as $v_{\rm phs}$ whereas that by locating the absorption trough as $v_{\rm pha}$.
  The phases are expressed in days with reference to the $t_{\rm ref}$ adopted in Table 1, while velocities are given in units of $10^{3} \kms$. Velocities at phases marked with astrisks are estimated using \Hei\ lines.

\end{table*}

\section{data} \label{sec:dat}

The sample of type IIP SNe consists of two subluminous SNe 2005cs and 2009md; two normal-luminosity
SNe 1999gi and 2012aw; three bright SNe 2004et, 2006bp and 2009bw and a intermediate luminosity SN 2008in having peculiar characteristics.
The basic properties of SNe and their host galaxies are given in Table~\ref{tab:par-sne}.
The time of explosion $t_{\rm ref}$ is determined from observational
non-detection in optical bands and it is constrained with an accuracy of a day for SNe 2005cs,
2004et and 2012aw, while for the remaining SNe, it is usually constrained by matching the
spectra with known template of IIP SNe and the accuracy lies between 2 to 8 days. The total interstellar
reddening $E(B-V)_{\rm tot}$ given in Table \ref{tab:par-sne} includes combined
reddening due to the Milky way and the host galaxy. For most of these SNe, the value of reddening
is constrained quite accurately. Moreover, in this work, values of extinction in different
filters (required as input in Eq.~7 and derived using adopted reddening) is estimated assuming
the line-of-sight ratio of
total-to-selective extinction Rv = 3.1, though a different reddening law towards the
sightline of highly embedded SNe cannot be ruled out. We study the implication of variation in
reddening on the distance determinations in \S\ref{sec:epm}.

The criterion for selecting the present sample has been the availability of photometric and spectroscopic
data on at least three phases by 50 days after explosion We restricted the use of data for \epm\ analysis
up to the phase 50d, as the value of $\xi$ depends on the color temperature and it varies sharply
below 5 kK, i.e. about 50d post explosion for IIP SNe. The $BVI$ photometric data are collected
from the literature and Fig.~\ref{fig:photplot_all} shows the photometric data used in this paper.
Barring SN 2009md, we have a dense coverage of early-time ($<50$ day) data for all the events.
A typical photometric accuracy for events brighter than 15 mag is 0.02 mag while for fainter
events it is poorer.

\begin{table*}[htbp]
\caption{EPM derived results for the events.}
\fontsize{2.7mm}{5mm}\selectfont
    \begin{tabular}{ll|ccc>{\bf}c|ccc>{\bf}c|}
    \cline{3-10}
          &       & \multicolumn{4}{|c|}{D05}      & \multicolumn{4}{|c|}{H01} \\
    \cline{3-10}
                                                        &       & $BV$             & $BVI$            & $VI$             & Mean           & $BV$             & $BVI$            & $VI$             & Mean            \\ \hline
    \multicolumn{1}{|c|}{\multirow{2}[0]{*}{SN 1999gi}} & $D$     & 11.92$\pm$1.08 & 11.34$\pm$0.34 & 11.60$\pm$0.86 & 11.62$\pm$0.29 &  8.64$\pm$0.80 &  8.71$\pm$0.31 &  9.27$\pm$0.73 &  8.87$\pm$0.34  \\
                                 \multicolumn{1}{|c|}{} & $t_{\rm 0}$ &  1.71$\pm$0.99 &  2.22$\pm$0.47 &  1.37$\pm$0.83 &  1.76$\pm$0.43 &  2.87$\pm$0.69 &  2.78$\pm$0.45 &  1.58$\pm$0.89 &  2.41$\pm$0.72  \\ \hline
    \multicolumn{1}{|c|}{\multirow{2}[0]{*}{SN 2004et}} & $D$     &  5.28$\pm$0.23 &  4.48$\pm$0.13 &  6.47$\pm$0.26 &  5.41$\pm$1.00 &  3.56$\pm$0.17 &  3.29$\pm$0.10 &  5.22$\pm$0.22 &  4.02$\pm$1.04  \\
                                 \multicolumn{1}{|c|}{} & $t_{\rm 0}$ &  0.28$\pm$0.86 &  4.64$\pm$0.53 &  0.95$\pm$0.94 &  1.96$\pm$2.35 &  2.39$\pm$0.75 &  5.88$\pm$0.56 &  1.72$\pm$0.93 &  3.33$\pm$2.23  \\ \hline
    \multicolumn{1}{|c|}{\multirow{2}[0]{*}{SN 2005cs}} & $D$     &  7.62$\pm$0.26 &  7.70$\pm$0.25 &  8.61$\pm$0.33 &  7.97$\pm$0.55 &  5.86$\pm$0.24 &  5.98$\pm$0.20 &  6.76$\pm$0.27 &  6.20$\pm$0.49  \\
                                 \multicolumn{1}{|c|}{} & $t_{\rm 0}$ & -0.49$\pm$0.68 & -0.35$\pm$0.54 & -1.77$\pm$0.73 & -0.87$\pm$0.78 &  0.14$\pm$0.73 &  0.18$\pm$0.51 & -1.24$\pm$0.73 & -0.31$\pm$0.81  \\ \hline
    \multicolumn{1}{|c|}{\multirow{2}[0]{*}{SN 2006bp}} & $D$     & 18.82$\pm$1.04 &       ---      &       ---      & 18.82$\pm$1.04 & 12.47$\pm$0.57 &       ---      &       ---      & 12.47$\pm$0.57  \\
                                 \multicolumn{1}{|c|}{} & $t_{\rm 0}$ & -3.23$\pm$0.77 &       ---      &       ---      & -3.23$\pm$0.77 & -0.95$\pm$0.50 &       ---      &       ---      & -0.95$\pm$0.50  \\ \hline
    \multicolumn{1}{|c|}{\multirow{2}[0]{*}{SN 2008in}} & $D$     & 13.11$\pm$0.68 & 14.56$\pm$0.76 & 15.86$\pm$0.83 & 14.51$\pm$1.38 & 12.71$\pm$0.84 & 11.62$\pm$0.64 & 12.58$\pm$0.64 & 12.31$\pm$0.59  \\
                                 \multicolumn{1}{|c|}{} & $t_{\rm 0}$ & -5.56$\pm$1.34 & -6.51$\pm$1.13 & -2.84$\pm$1.11 & -4.97$\pm$1.91 &-11.45$\pm$2.04 & -7.01$\pm$1.19 & -2.14$\pm$0.96 & -6.87$\pm$4.66  \\ \hline
    \multicolumn{1}{|c|}{\multirow{2}[0]{*}{SN 2009bw}} & $D$     & 15.70$\pm$1.67 & 16.15$\pm$1.07 & 22.26$\pm$1.57$^*$& 15.93$\pm$0.32 & 12.53$\pm$1.39 & 12.68$\pm$0.98 & 17.11$\pm$1.36$^*$ & 12.61$\pm$0.11 \\
                                 \multicolumn{1}{|c|}{} & $t_{\rm 0}$ & -0.49$\pm$4.90 & -2.27$\pm$1.87 &-12.28$\pm$5.69$^*$ & -1.38$\pm$1.26 & -2.55$\pm$8.10 & -2.46$\pm$2.12 &-10.35$\pm$4.25$^*$ & -2.51$\pm$0.06 \\ \hline
	\multicolumn{1}{|c|}{\multirow{2}[0]{*}{SN 2009md}} & $D$     & 21.06$\pm$4.21 & 24.09$\pm$3.79 & 24.72$\pm$3.62 & 23.29$\pm$1.96 & 18.74$\pm$3.44 & 20.29$\pm$3.24 & 19.29$\pm$2.43 & 19.44$\pm$0.78  \\
                                 \multicolumn{1}{|c|}{} & $t_{\rm 0}$ &  6.29$\pm$0.46 &  2.42$\pm$4.78 &  3.73$\pm$4.00 &  4.15$\pm$1.97 &  2.23$\pm$0.46 &  0.16$\pm$5.18 &  4.55$\pm$3.16 &  2.31$\pm$2.19  \\ \hline
	\multicolumn{1}{|c|}{\multirow{2}[0]{*}{SN 2012aw}} & $D$     & 11.06$\pm$0.44 & 10.51$\pm$0.21 & 12.24$\pm$0.49 & 11.27$\pm$0.88 &  8.22$\pm$0.35 &  8.06$\pm$0.16 &  9.72$\pm$0.43 &  8.67$\pm$0.92  \\
                                 \multicolumn{1}{|c|}{} & $t_{\rm 0}$ & -2.55$\pm$0.71 & -1.53$\pm$0.32 & -2.99$\pm$0.71 & -2.36$\pm$0.75 & -1.74$\pm$0.61 & -0.86$\pm$0.35 & -2.44$\pm$0.82 & -1.68$\pm$0.79  \\ \hline \hline

	\multicolumn{10}{|c|}{EPM with fixed explosion epoch} \\ \hline
	\multicolumn{1}{|c|}{\multirow{1}[0]{*}{SN 2004et}} & $D$     &  5.36$\pm$0.13 &  5.50$\pm$0.05 &  6.73$\pm$0.10 &  5.86$\pm$0.76 &  4.07$\pm$0.09 &  4.32$\pm$0.04 &  5.60$\pm$0.07 &  4.66$\pm$0.82 \\ \hline
	\multicolumn{1}{|c|}{\multirow{1}[0]{*}{SN 2005cs}} & $D$     &  7.34$\pm$0.19 &  7.52$\pm$0.18 &  7.62$\pm$0.19 &  7.49$\pm$0.14 &  5.93$\pm$0.16 &  6.05$\pm$0.13 &  6.19$\pm$0.15 &  6.06$\pm$0.13 \\ \hline
    \multicolumn{1}{|c|}{\multirow{1}[0]{*}{SN 2012aw}} & $D$     &  9.46$\pm$0.27 &  9.74$\pm$0.12 & 10.27$\pm$0.27 &  9.83$\pm$0.41 &  7.30$\pm$0.20 &  7.70$\pm$0.09 &  8.41$\pm$0.22 &  7.80$\pm$0.56 \\

    \hline
    \end{tabular}
  \label{tab:dis-epm}
  \fontsize{2.7mm}{3.5mm}\selectfont
  \newline
   Notes: $D$ denotes the distance in Mpc. $t_0$ denotes the time of explosion in days, derived in this study and measured with reference to the adopted time of explosion ($t_{\rm ref}$) in Table~\ref{tab:par-sne}.
   Negative values of $t_0$ indicate dates prior to the adopted value. The values marked with asterisks are considered deviant and these are not considered in computing the mean value.
\end{table*}

We obtained the wavelength- and flux-calibrated spectra either from
SUSPECT\footnote{http://suspect.nhn.ou.edu/$ \sim $suspect/} database or from corresponding
authors of papers (see Table \ref{tab:par-sne}). A typical spectral resolution in the visible range of
spectra lies between 5 to 10\AA\ ($\sim$ 300 to 600 \kms at 5500 \AA). For SN 2004et, we have also
included 6 epoch spectra between +11d to +16d from \cite{takats12}. The spectra were corrected for
respective recession velocity of their host galaxy before estimating the photospheric velocity.
Table~\ref{tab:vel-sne} provides value of photospheric velocities derived using both the
methods described in \S\ref{sec:vel_det}, i.e. $v_{\rm pha}$ and $v_{\rm phs}$. A detailed
description of the \synow\ modelling of spectra and determination of $v_{\rm phs}$ and its error
followed in this work is given elsewhere \citep{bose13aw}. We briefly describe the method below.
As we are only interested in obtaining photospheric velocity we fit the observed and synthetic spectra
locally around \Feii\ lines (4923.93, 5018.44 and 5169.03 \AA) within 4700 - 5300\AA, and
in early phases where \Feii\ lines are not available we fit around \Hei\ 5876\AA\ line within 5500-6200 \AA\ only; since employing the whole
wavelength range may introduce over or under-estimation of photospheric velocities as different lines form
at different layers. After attaining optimal fit of observed spectra locally, we only vary
model parameter $v_{\rm ph}$ to get eye estimate of maximum possible deviation from optimal value and this
is attributed as the uncertainty in $v_{\rm ph}$ for that spectrum. We note that as P-Cygni profiles are
quite sensitive to $v_{\rm ph}$ and hence the best fits are easily attainable through eye inspection.
The typical uncertainty in velocities estimated by deviation seen visually from best-fit
absorption troughs varies between 50 to 500 \kms with a typical value of $\sim$~150 \kms.
This is consistent with the values obtained using automated computational techniques
viz. $\chi^{2}$-minimization and cross-correlation methods employing entire spectra \citep{takats12}.
A comparison of $v_{\rm pha}$ and $v_{\rm phs}$ is also made and deviations as large as
1000 \kms\, is seen in early spectra for a a few SNe, while random deviations are apparent at later epochs
to the level of quoted uncertainty. We study implication of using these velocities on the
distance determinations in \S\ref{sec:epm}.

\begin{figure}
\centering
\includegraphics[width=7cm]{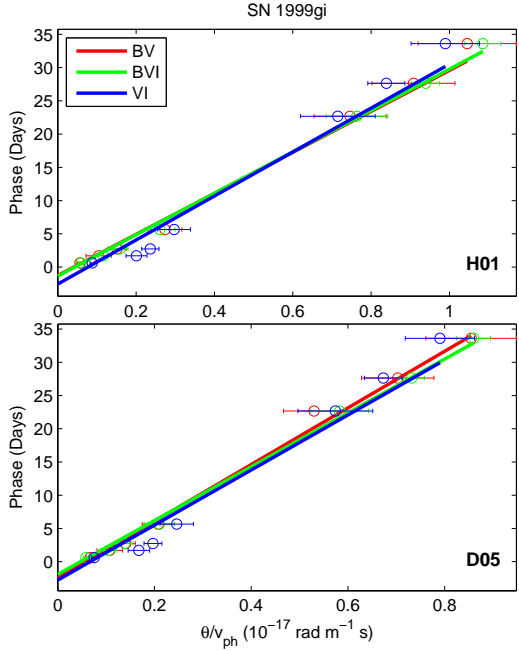}
\caption{\epm\, fitting for SN 1999gi using two sets of dilution factors
 H01 (top) and D05 (bottom) in combination to three filter subsets BV, BVI and VI. The phases are in
 reference to the corresponding $ t_{\rm ref} $ adopted in table \ref{tab:par-sne}.}
\label{fig:epm_99gi}
\end{figure}

\begin{figure}
\centering
\includegraphics[width=7cm]{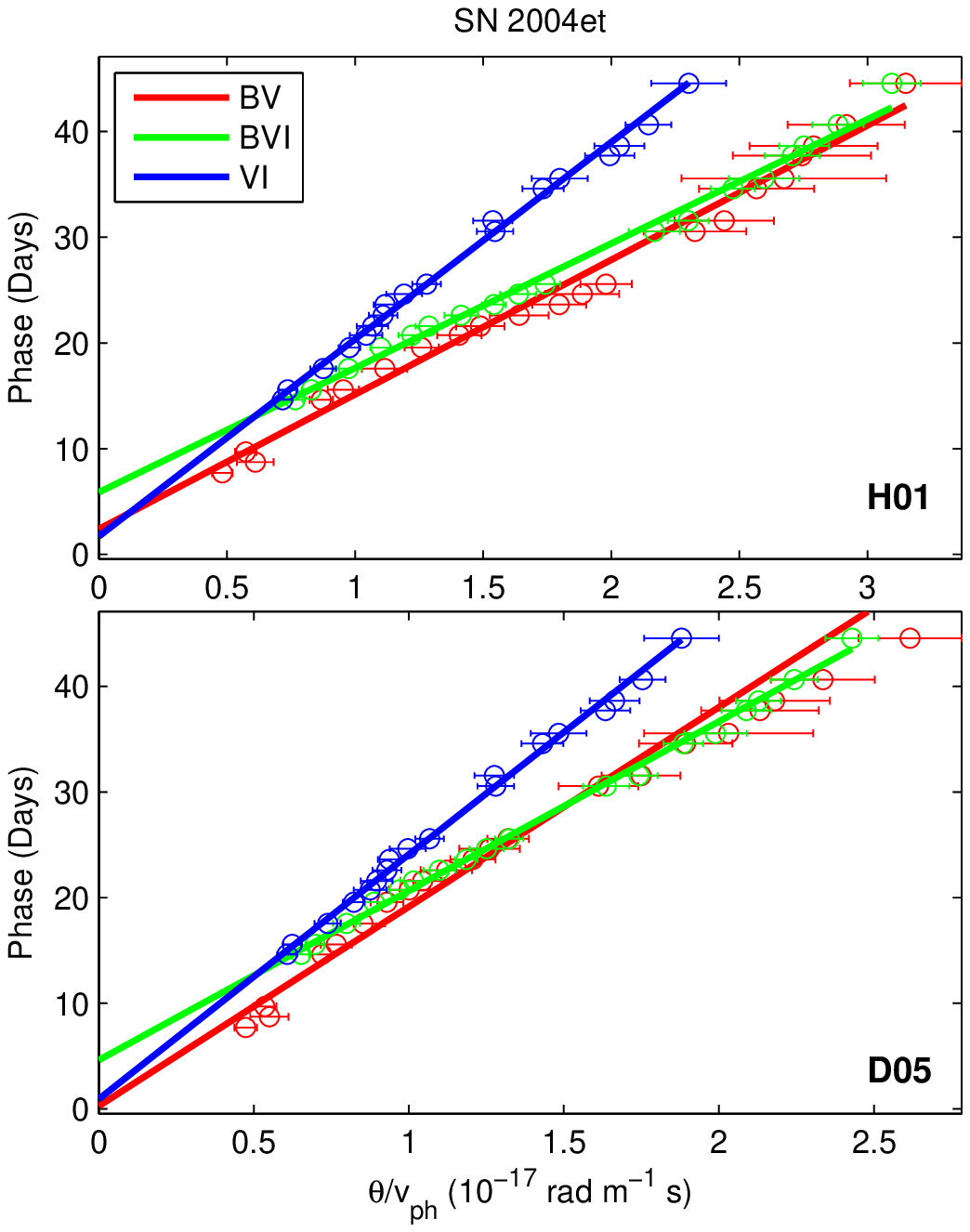}
\caption{\figcap{2004et}{2453270.5}}
\label{fig:epm_04et}
\end{figure}

\begin{figure}
\centering
\includegraphics[width=7cm]{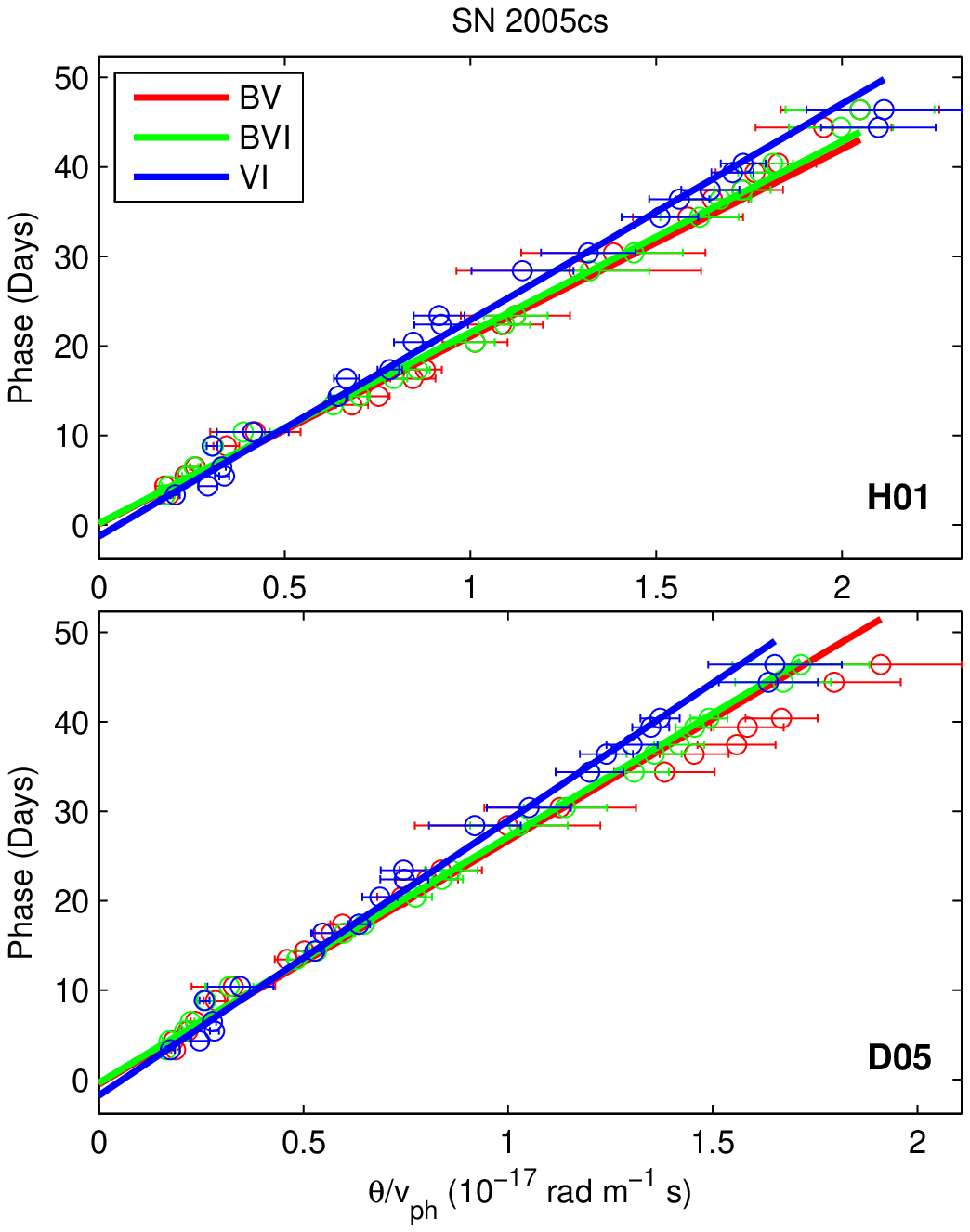}
\caption{\figcap{2005cs}{2453549.0}}
\label{fig:epm_05cs}
\end{figure}

\begin{figure}
\centering
\includegraphics[width=7cm]{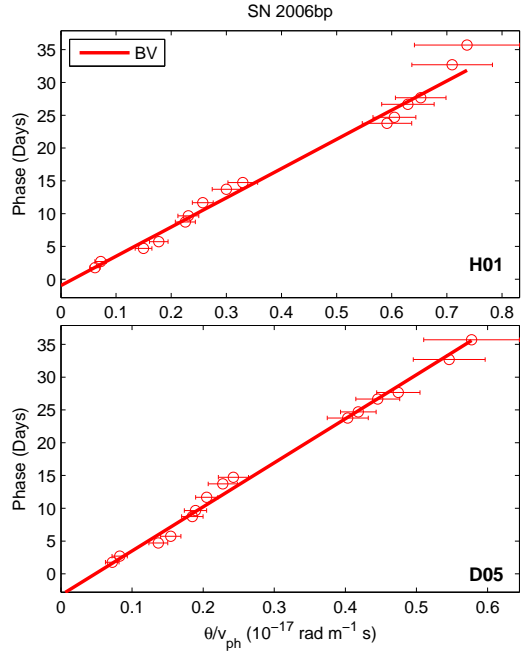}
\caption{
Same as Fig. \ref{fig:epm_99gi}, but for SN 2006bp using only BV filter subset.
}
\label{fig:epm_06bp}
\end{figure}

\begin{figure}
\centering
\includegraphics[width=7cm]{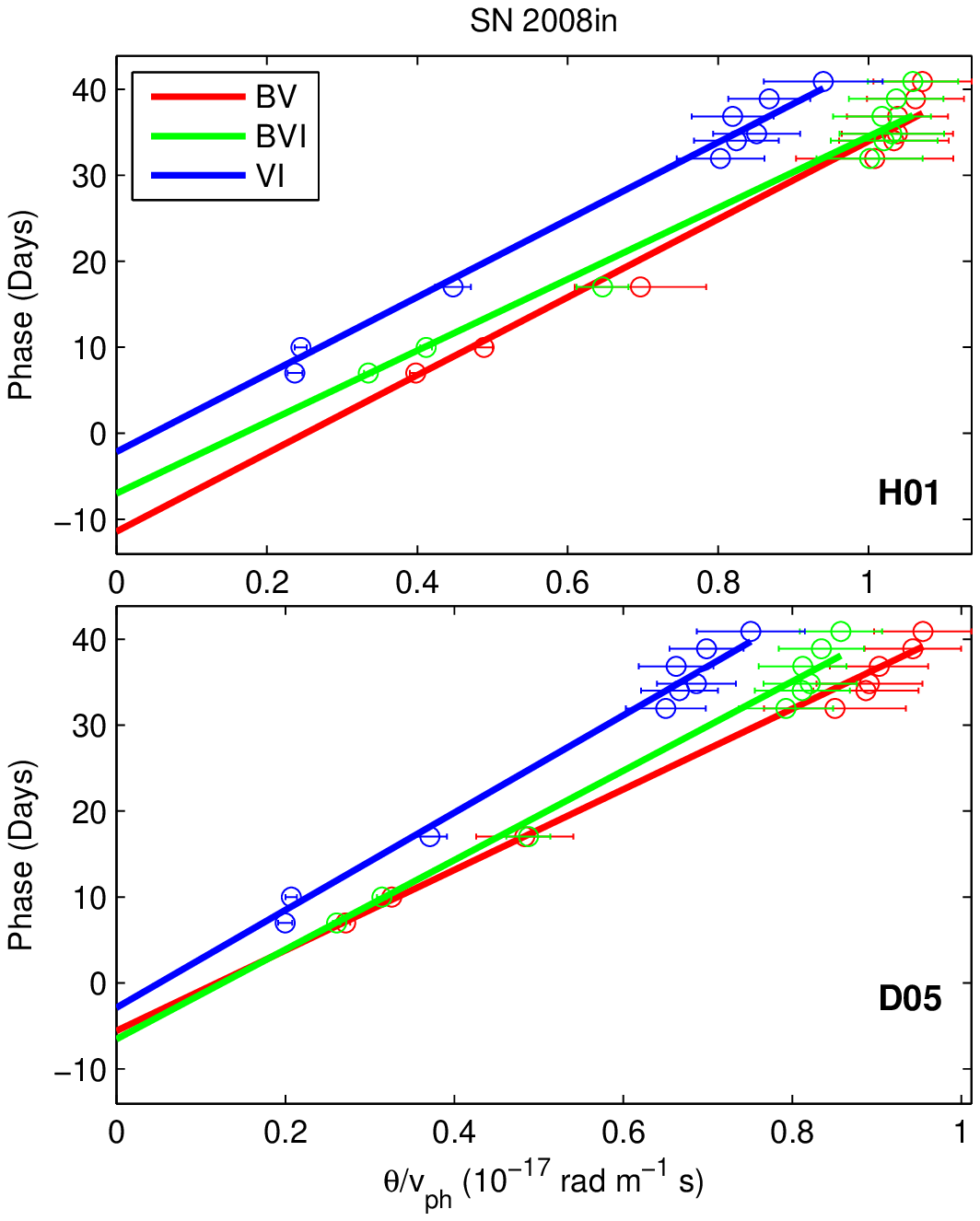}
\caption{\figcap{2008in}{2454825.6}}
\label{fig:epm_08in}
\end{figure}

\begin{figure}
\centering
\includegraphics[width=7cm]{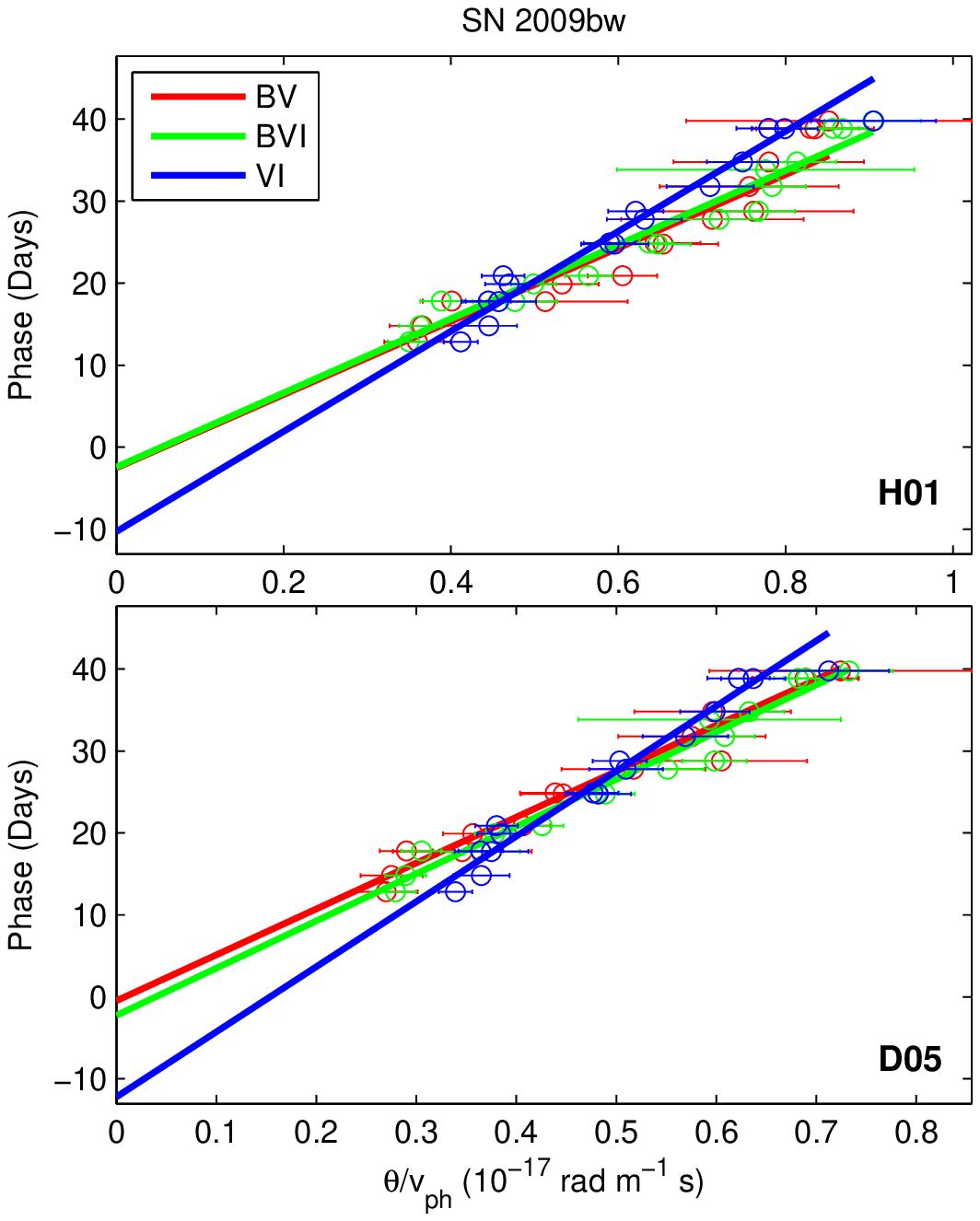}
\caption{\figcap{2009bw}{2454825.6}}
\label{fig:epm_09bw}
\end{figure}

\begin{figure}
\centering
\includegraphics[width=7cm]{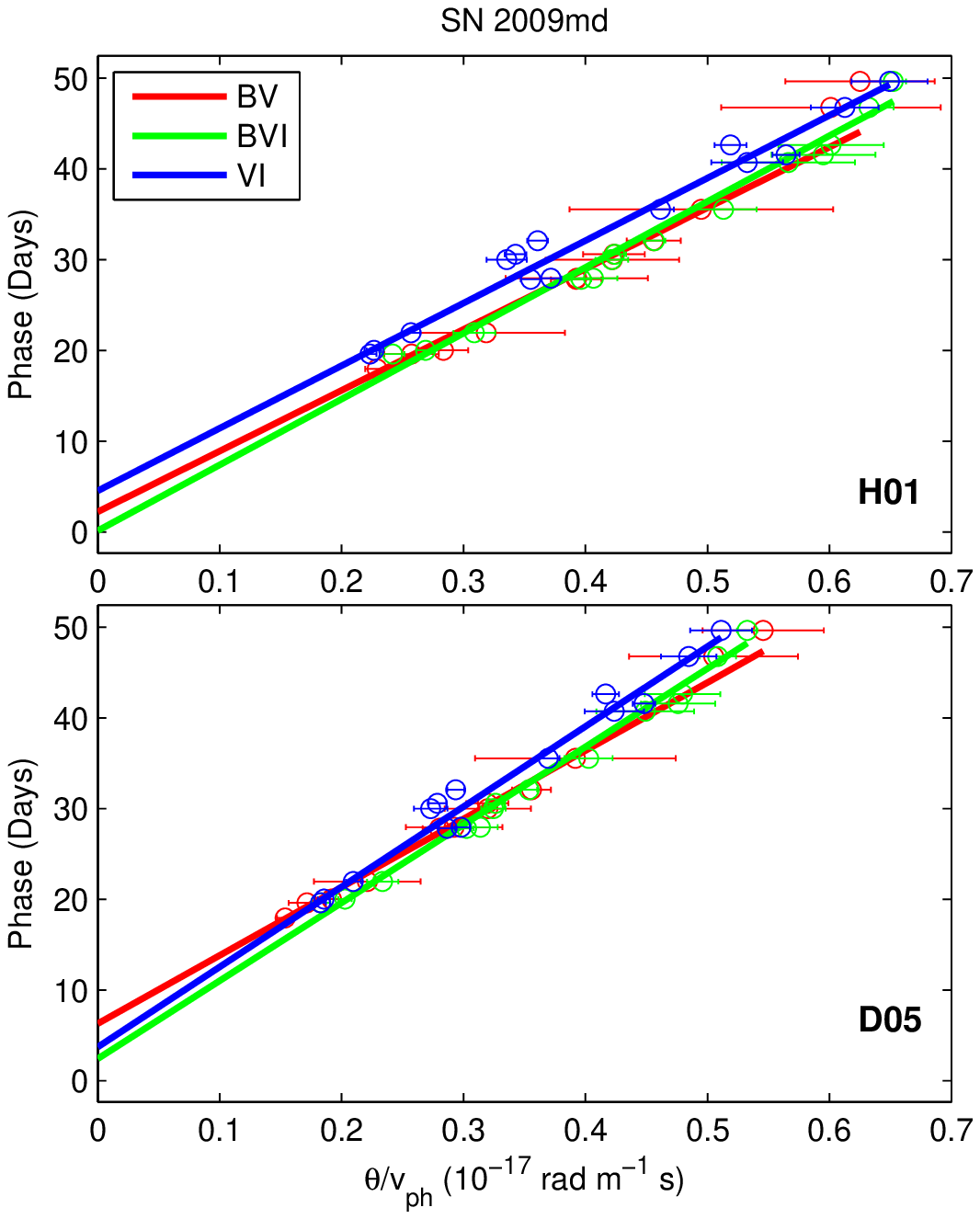}
\caption{\figcap{2009md}{2455162.0} Error bars are reduced by factor of five.}
\label{fig:epm_09md}
\end{figure}

\begin{figure}
\centering
\includegraphics[width=7cm]{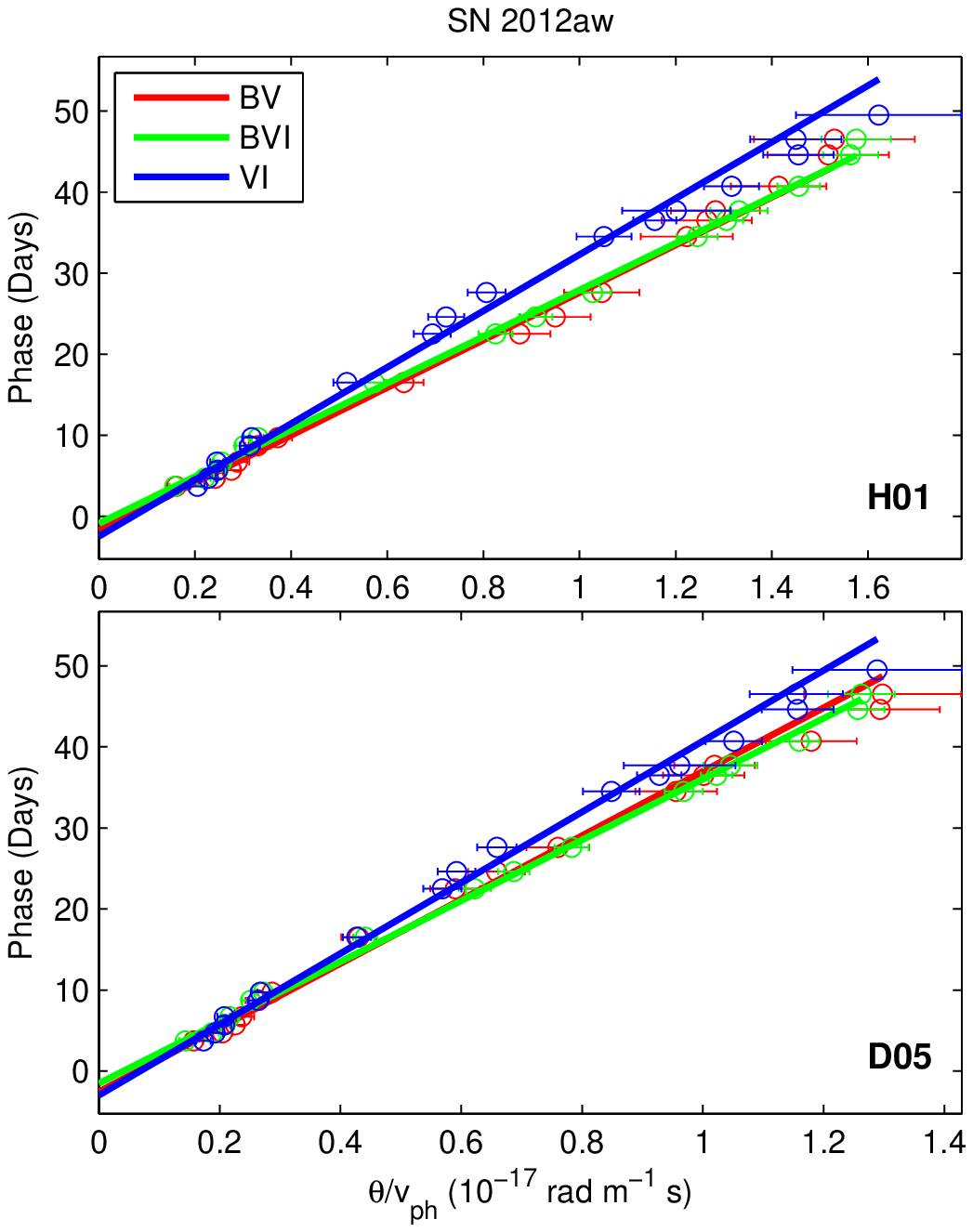}
\caption{\figcap{2012aw}{2456002.6}}
\label{fig:epm_12aw}
\end{figure}

\begin{figure}
\centering
\includegraphics[width=7cm]{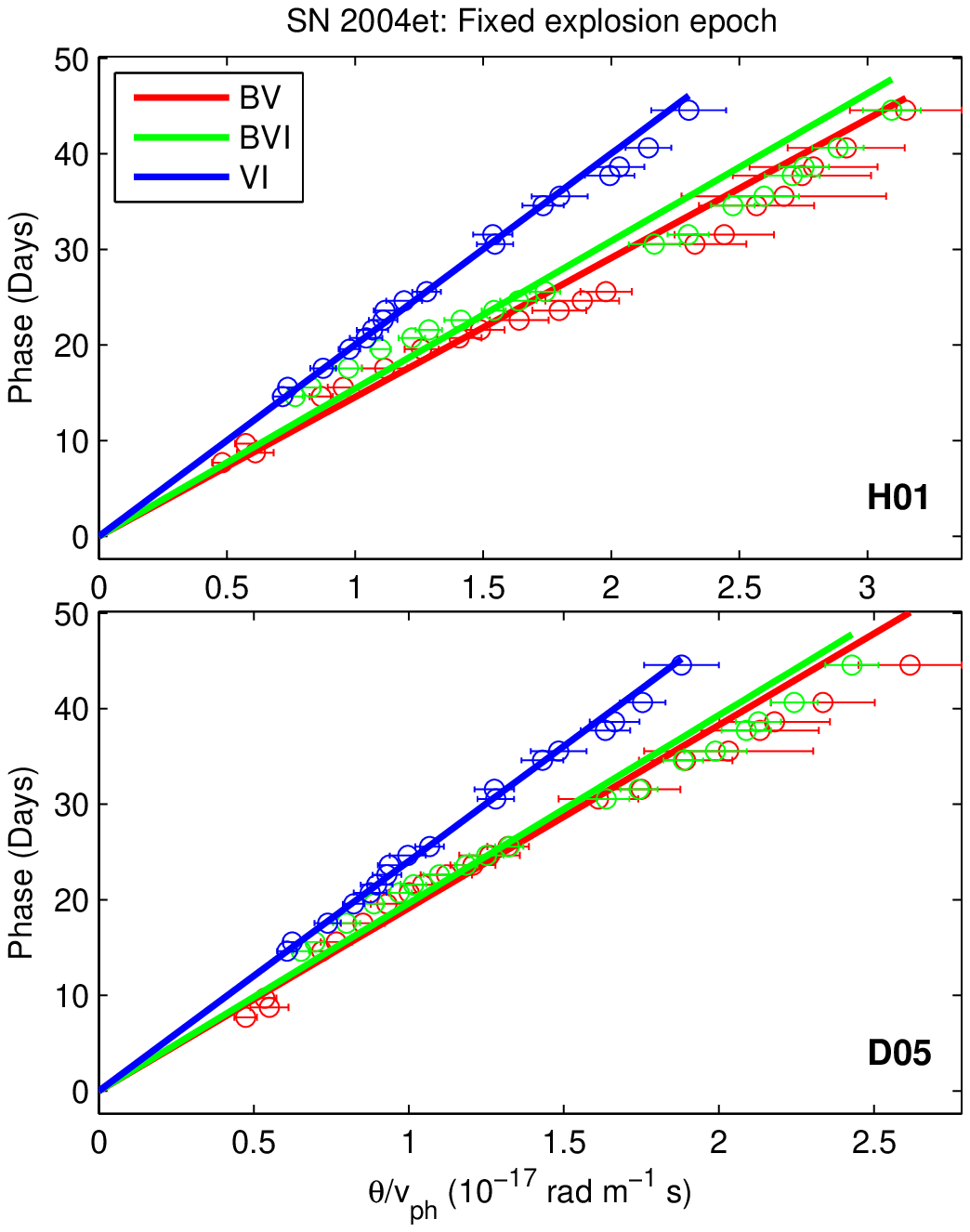}
\caption{\figcapFix{2004et}{2453270.5}}
\label{fig:epm_04et_fixt0}
\end{figure}

\begin{figure}
\centering
\includegraphics[width=7cm]{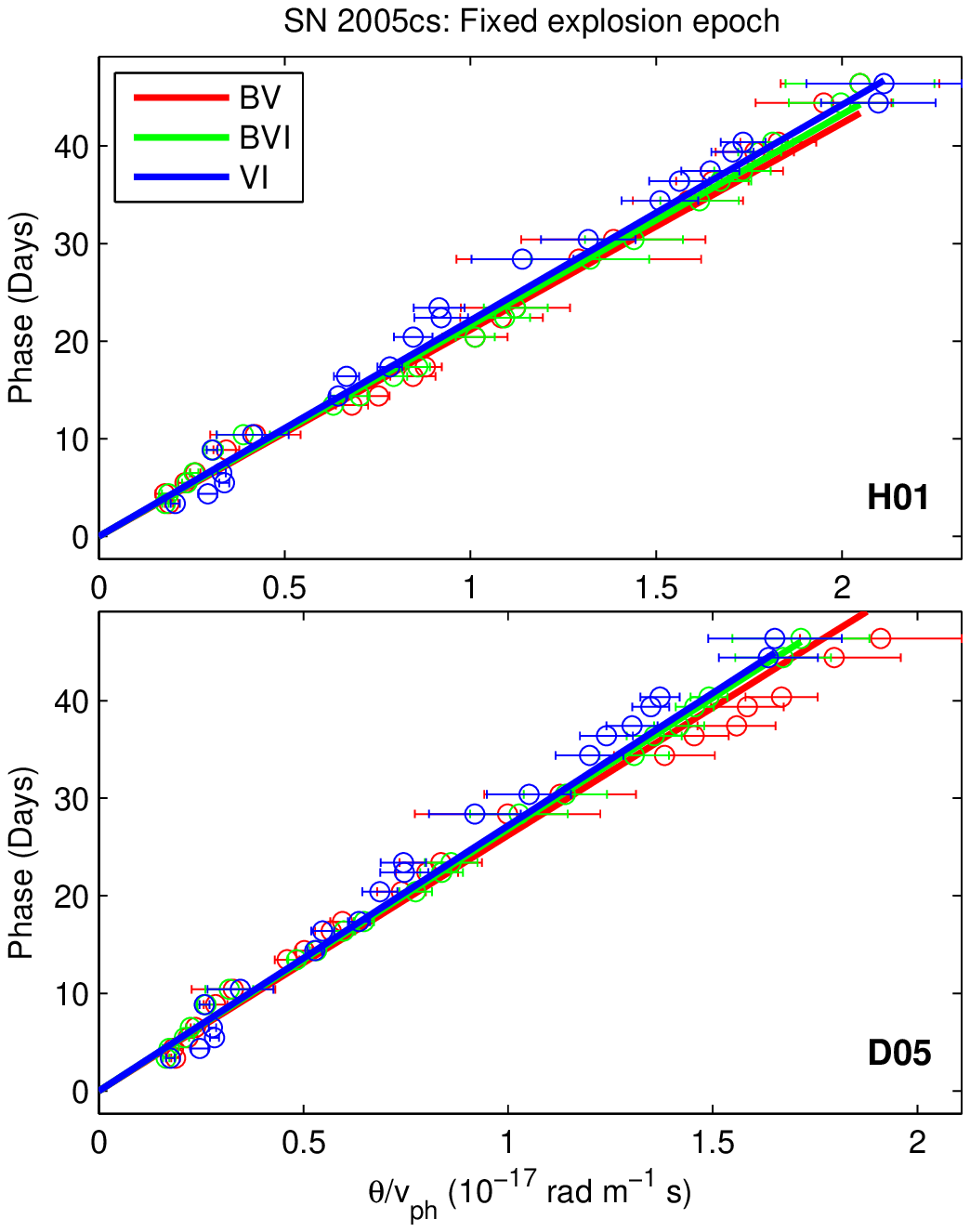}
\caption{\figcapFix{2005cs}{2453549.0}}
\label{fig:epm_05cs_fixt0}
\end{figure}

\begin{figure}
\centering
\includegraphics[width=7cm]{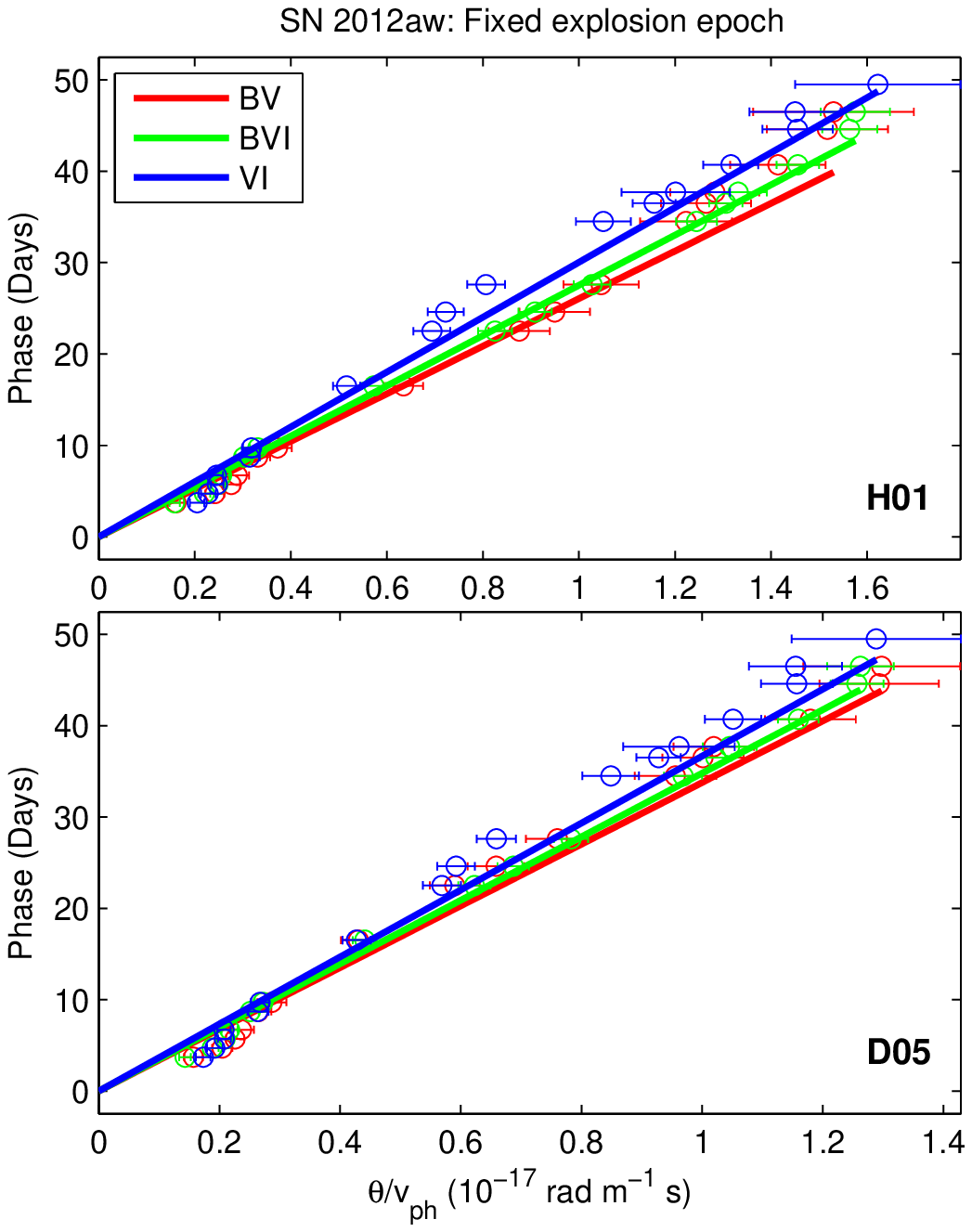}
\caption{\figcapFix{2012aw}{2456002.6}}
\label{fig:epm_12aw_fixt0}
\end{figure}

\section{EPM analysis} \label{sec:epm}

At each $t$ for which photometric data is available, we derive the value of $\theta$ for three sets of
filter combinations and for two sets of $\xi$ prescriptions. Wherever spectroscopic data do not
coincide with the epoch of photometric data point, the value of $v_{\rm ph}$ at $t$ is derived by
polynomial interpolation of third or fourth order. It is noted here that in comparison to photometry, spectroscopy of SNe
is more demanding in terms of telescope time and as a result, many of our
sample have large spectroscopic data gap. In this work, we have therefore, opted to use the interpolated
spectroscopic data for the corresponding epoch of photometric data presented in Fig.~\ref{fig:photplot_all}.
We performed $\chi^{2}$-minimization for $\theta/v_{\rm ph}$ versus
time to derive $D$ and $t_{0}$ (see Eq~\ref{eq:epm}). Here, we use \synow-derived value of $v_{\rm ph}$
(i.e. $v_{\rm phs}$, see Table~\ref{tab:vel-sne}).
Fig.~\ref{fig:epm_99gi} to~\ref{fig:epm_12aw} show plots for SNe 1999gi, 2004et, 2005cs, 2006bp,
2008in, 2009bw, 2009md, and 2012aw respectively whereas Fig.~\ref{fig:epm_04et_fixt0}
to~\ref{fig:epm_12aw_fixt0} show plots to estimate $D$ with fixed $t_{0}$ ($=t_{\rm ref}$, see
Table~\ref{tab:par-sne}) for SNe 2004et, 2005cs and 2012aw.

The results are listed in Table~\ref{tab:dis-epm}. The errors quoted for distance and explosion epoch
are mainly contributed by errors in $\theta$ and $v_{\rm ph}$; we discuss errors briefly.
Error in quantities $\theta \xi$ and $T$ (see Eq.~\ref{eq:chisq}) for a fixed value of \ebv\ are
estimated using Monte Carlo technique in which a sample of one thousand data points are drawn from normal distribution of uncertainty in the observed photometric fluxes. Considering
that $\xi$ is one dimensional function of temperature only, the error in $\xi$ is numerically estimated
using error in $T$. So, the error in $\theta$ is computed by combining errors of $\xi$ and $\theta\xi$ in
quadrature. We note that, intrinsically, the factor $\xi$ is a major source of systematic error and it may
lead to over or under estimation of \epm\--derived distance.

The source of error in $v_{ph}$ is random in nature and the relative error in it varies
between 2 to 5\% whereas in $\theta$ it varies between 5 to 10\%. While interpolating velocities at
desired photometric epochs, the errors are estimated by Monte Carlo method with a sample size of 1000.
For the final \epm\ fit, the error in $\theta/v_{ph}$ is propagated from $\theta$ and $v_{ph}$ and
the weighted least-squared fitting is performed to estimate distance and explosion epoch. The error in finally
derived distance for each filter subset is estimated by Monte Carlo technique with a sample size of 1000.

It can be seen from Table~\ref{tab:dis-epm} that for each prescription, we derive three sets of $D$
and $t_{0}$ corresponding to
each of the three filter sets. Barring SN 2009bw, the values of $D$ and $t_{0}$ for each of the filter sets
are consistent within uncertainties. We, therefore, combine individual distances and explosion epochs
derived for each filter set, to compute mean values of $D$ and $t_{0}$ for D05 and H01 $\xi$ prescriptions.
The quoted uncertainty in the mean values is the standard deviation of the values obtained for the three
filter sets and it can be seen that statistical errors in mean value are consistent with the errors
derived in individual filter-sets, barring the case of SN 2009bw which has deviant values for $VI$ set.
It can be noted that the relative precision with which \epm\, distances are derived for either of the
atmosphere models (D05 or H01) lies between 2 to 13\% having a median value of 6\%.

\begin{figure}
\centering
\includegraphics[width=8cm]{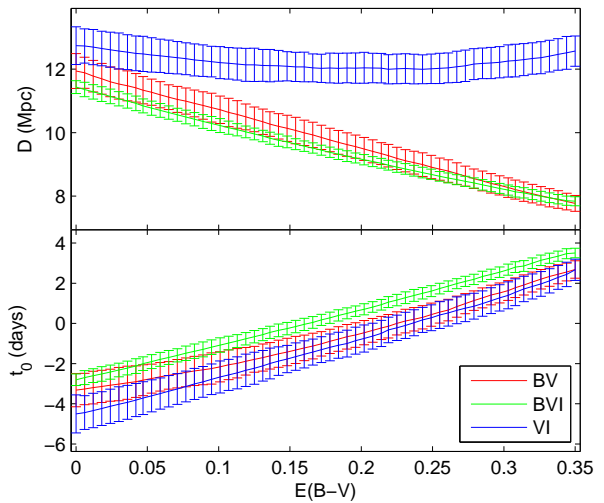}
\caption{Variation of \epm\ distances and explosion epoch for each filter sub-set \{BV\}, \{BVI\} and \{VI\}
                with the variation of E(B-V) for the SN 2012aw using D05 prescription.}
\label{fig:aw_ebv_var_d05}
\end{figure}

Another source of error in $D$ and $t_0$ is the value of \ebv. Though, we have taken its value from literature,
and adopted value derived using most reliable method, but its precise determination is extremely
difficult and it can introduce systematic error in determination of \epm\ distance.  We have studied
the effect of \ebv\ for SN 2012aw by varying its
value for each filter subset. Figure \ref{fig:aw_ebv_var_d05} shows the variation of \epm\ distance
and explosion epoch with \ebv. The variation of distance differs significantly among each filter
subset, however the overall variation in distance is not very significant.  In order to further study the
effect of \ebv\, variation on \epm\, results for each SNe, we derive mean distances and explosion
epochs with the upper and lower limit of \ebv\ and tabulate them in Table~\ref{tab:ebv-err}. We took this approach
to estimate deviations of \epm\ results from corresponding \ebv\ errors because of it's systematic dependence
on analysis and we found that it would have been inappropriate to propagate \ebv\ errors all throughout the
analysis. It is noted here that \epm\ results have non-linearly dependence on \ebv\ and thus the resulting tabulated
errors are asymmetric. The relative variation in $ D $ is found to lie between 0 to 9\% with median value of 5\%.

\begin{table}
  \caption{Dependence of \epm\, derived parameters on the errors of \ebv.}
  \label{tab:ebv-err}
\fontsize{2.8mm}{5mm}\selectfont
  \begin{tabular}{l c c c} \hline
    SN event&    $E(B-V)$        &$D$                                         &  $t_0$ \\
			&(mag)		         & (Mpc) 			                          &  (day) \\ \hline
	SN 1999gi &  \ve{0.21}{0.09} & $ 11.62^{-0.78\pm0.50}_{+0.93\pm0.51} $ & $  ~~1.76^{+0.95\pm0.49}_{-0.87\pm0.46} $ \\
	SN 2004et &  \ve{0.41}{0.04} & $ ~5.86^{+0.09\pm0.96}_{-0.05\pm0.58} $ & $               -                      $ \\
	SN 2005cs &  \ve{0.05}{0.02} & $ ~7.97^{-0.20\pm0.62}_{+0.11\pm0.56} $ & $  -0.87^{+0.37\pm0.84}_{-0.22\pm0.93} $ \\
	SN 2006bp &  \ve{0.40}{0.04} & $ 18.82^{-1.10\pm1.12}_{+0.99\pm1.04} $ & $  -3.23^{+0.94\pm0.89}_{-0.72\pm0.77} $ \\
	SN 2008in &  \ve{0.10}{0.10} & $ 14.51^{-1.13\pm1.93}_{+1.34\pm1.81} $ & $  -4.97^{+1.68\pm2.18}_{-1.50\pm1.41} $ \\
	SN 2009bw &  \ve{0.31}{0.03} & $ 15.93^{-0.62\pm0.09}_{+0.79\pm0.50} $ & $  -1.38^{+0.69\pm0.57}_{-0.92\pm1.71} $ \\
	SN 2009md &  \ve{0.10}{0.05} & $ 23.29^{-0.92\pm1.88}_{+1.10\pm2.04} $ & $ ~~4.15^{+0.56\pm1.58}_{-0.62\pm2.27} $ \\
    SN 2012aw &  \ve{0.07}{0.01} & $ ~9.83^{-0.02\pm0.43}_{+0.00\pm0.30} $ & $               -                      $ \\ \hline

  \end{tabular}
  \footnotesize
  \newline
  The superscript and subscript values in $ t_0 $ and $ D $ signify the values derived using upper and lower value of \ebv\ respectively. Further uncertainties quoted in these values are the standard deviation of values obtained from three band sets for each limit of \ebv.

  The references for the values and corresponding errors of \ebv\, are given in Table~\ref{tab:par-sne}. The errors for SN 2004et and SN 2006bp were unavailable in literature, thus for sake of reasonable approximation, we have attributed 10\% error in \ebv\ for these SNe.

\end{table}

Table~\ref{tab:dis-ned} compares mean value of distances to host galaxies which are taken from \ned\, and
are derived using redshift independent methods, such as Cepheids, Tully-Fisher, Standard Candle Method, Surface Brightness Fluctuation. with that derived
using \epm\,. The
comparison clearly illustrates that the distance derived using D05 prescription is in better agreement with the
\ned\, ones, whereas the ones using H01 prescription are systematically lower in each of
the cases. Similar systematic differences in the two atmosphere models (D05,H01) have also been
reported in the EPM implementation to 12 type IIP SNe by \cite{2009ApJ...696.1176J}.

For D05 models, a comparison of distances derived using $v_{\rm phs}$ and $v_{\rm pha}$,
see Table~\ref{tab:dis-phs}, indicate that barring a few cases, there is notable difference in both of the
value of distances. For SN 2005cs and 2009md the difference is as high as 18 and 15\% respectively,
for SN 1999gi, 2004et and 2009bw the values differ by 6 - 9\%. However, for SN 2006bp, 2008in and 2012aw the
difference is quite negligible and lies in 0 - 3\% which is within the internal precision of both values.

EPM analysis of the individual cases are discussed in \S\ref{sec:dis}.

\begin{table}
\centering
  \caption{ Comparison of \epm\, distances to host galaxies with that derived using other methods.}
  \label{tab:dis-ned}

  \begin{tabular}{l l c c} \hline
    Host    &Supernova  &$ D_{\epm} $& $ D_{\ned} $  \\
    Galaxy  &Event           &(Mpc)             &(Mpc)  \\ \hline
	NGC 3184& SN 1999gi&  11.62$\pm$0.29 &11.95$\pm$2.71 \\
	NGC 6946& SN 2004et&  ~5.86$\pm$0.76 &~5.96$\pm$1.97 \\
	NGC 5194& SN 2005cs&  ~7.97$\pm$0.55 &~7.91$\pm$0.87 \\
	NGC 3953& SN 2006bp&  18.82$\pm$1.04 &18.45$\pm$1.60 \\
	NGC 4303& SN 2008in&  14.51$\pm$1.38 &16.46$\pm$10.8 \\
	UGC 2890& SN 2009bw&  15.93$\pm$0.32 &      ---      \\
	NGC 3389& SN 2009md&  23.29$\pm$1.96 &21.29$\pm$2.21 \\
        NGC 3351& SN 2012aw&  ~9.83$\pm$0.41 &10.11$\pm$0.98 \\ \hline
  \end{tabular}
  \flushleft
  Notes: $D_{\ned}$ denote distances to host galaxies as collected from \ned\, (http://ned.ipac.caltech.edu)
         and derived using redshift-independent methods (see \S\ref{sec:epm}). $D_{\epm}$,
         taken from Table~\ref{tab:dis-epm}, denote \epm\, distances derived using D05 atmosphere
         model and \synow\ derived
         velocities $v_{\rm phs}$. For SNe 2004et and 2012aw, the distance are with fixed $ t_0 $.

\end{table}

\begin{table}
\centering
  \caption{Comparison of EPM distances derived using \synow\ modeled velocities and from velocities determined by locating the absorption trough.}
  \label{tab:dis-phs}

  \begin{tabular}{l c c} \hline
    SN event&$D_{\rm phs}$&$D_{\rm pha}$ \\
			&(Mpc)		 & (Mpc) 			\\ \hline
	SN 1999gi &  11.62$\pm$0.29 &   12.72$\pm$0.47  \\
	SN 2004et &  ~5.41$\pm$1.00 &   ~4.96$\pm$0.88  \\
	SN 2005cs &  ~7.97$\pm$0.55 &   ~6.56$\pm$0.42  \\
	SN 2006bp &  18.82$\pm$1.04 &   18.13$\pm$1.18  \\
	SN 2008in &  14.51$\pm$1.38 &   14.52$\pm$1.39  \\
	SN 2009bw &  15.93$\pm$0.32 &   16.92$\pm$0.62  \\
	SN 2009md &  23.29$\pm$1.96 &   26.84$\pm$3.78  \\
    SN 2012aw &  11.27$\pm$0.88 &   11.27$\pm$0.92  \\ \hline

  \end{tabular}
  \newline
\begin{flushleft}
  $D_{\rm phs}$  denote \epm\ distances derived using \synow\ model velocities, i.e.  $v_{\rm phs}$, whereas,
  $D_{\rm pha}$  denote the ones derived by locating the absporption minima
  of \Feii\ lines, i.e. $v_{\rm pha}$. For consistency, the D05 presciption and unconstrained explosion epoch
  have been used for all the cases.
\end{flushleft}

\end{table}

\section{discussions} \label{sec:dis}

In the following, we shall discuss results for each of the event and also any anomaly if
found in the result.

SN 1999gi :
The photometric and spectroscopic data are taken from \cite{leon02} and the epm-fitting is shown
in Fig.~\ref{fig:epm_99gi}. For H01 $\xi$ prescription, we derived a
distance of $8.87\pm0.34$ Mpc whereas, \cite{leon02} and \cite{2009ApJ...696.1176J} derived a
value of \ve{11.1}{2.0} Mpc and \ve{11.7}{0.8} Mpc respectively. We attribute a lower value of
distance in our case to the method adopted in this work i.e. velocity estimates using \synow\
which is significantly different in some epochs and the filter-response deconvolution in SED fitting;
and also to less number of data points available for SN 1999gi. Removing first photometric point,
our EPM implementation yields a value of $\sim$ 10 Mpc.

For D05 prescription, we derived a value of $11.62\pm0.29$ Mpc while \cite{2009ApJ...696.1176J} derived a value of
\ve{17.4}{2.3} Mpc. It is noted that later has excluded the first spectroscopic data point and have used the
spectroscopic epochs for \epm\ fitting, in contrast to photometric epochs used in the present work,
see \S\ref{sec:epm}. Our estimate for D05 is in good agreement with the other redshift independent
estimate, see Tab~\ref{tab:dis-ned}.

SN 2004et :
We used 21 epochs of photometric data taken from \cite{sahu06} and \cite{takats12} to derive \epm\ distance. For this event
the time of explosion is determined observationally with an accuracy of a day, and hence, \epm\ fitting is
attempted and shown with $ t_0 $ as free and fixed parameters respectively in Fig.~\ref{fig:epm_04et}
and Fig.~\ref{fig:epm_04et_fixt0}. For D05 prescription, we derive a \epm\ distance of \ve{5.41}{1.00} Mpc
and \ve{5.86}{0.76} Mpc respectively; which are consistent with each other as
well as with the host galaxy distances derived using other methods. For the former $t_0$ is estimated
as \ve{1.96}{2.35} days; which is also consistent with the time of explosion adopted from
literature ($ t_{\rm ref} $). \cite{takats12} derived an \epm\
distance of $4.8\pm0.6$ using D05 prescriptions and \synow\ velocities.

However, it is noted that the $ VI $ fit is quite inconsistent in comparison to the $ BV $ and $ BVI $ sets.
In order to understand this discrepancy, we looked into the possibility of lower value of \ebv. \cite{sahu06}
stated that they found equivalent width of 1.7\AA\ for \ion{Na}{I}D from low resolution spectra which corresponds to
total \ebv$ =0.43$ mag, employing empirical relation of \cite{1990A&A...237...79B}. On adopting a similar empirical relation
by \cite{2003fthp.conf..200T} which we find more convincing, we arrive at a much lower value of \ebv\ which is 0.26 mag.
Being backed by this possibility of lower \ebv, we re-derive \epm\ distances considering only the Galactic reddening value
of 0.29 mag \citep{2011ApJ...737..103S} and arrive to \epm\ distances 5.59, 5.65 and 6.14 Mpc for $ BV $, $ BVI $ and
$ VI $ band sets respectively which are fairly consistent with each other. Despite of favorable results with lower \ebv\, we can not
rule out the higher value of \ebv$ =0.40 $ mag which was derived by \cite{2004IAUC.8413....1Z} using high resolution
spectra.

SN 2005cs:
We have used 14 epochs spectroscopic and 22 epoch of photometric data from \cite{pasto06,pasto09}.
This is another event for which explosion epoch is constrained observationally within a day
hence we have done the fitting by keeping $ t_0 $ as free (Fig.~\ref{fig:epm_05cs}) as well as
fixed (Fig. \ref{fig:epm_05cs_fixt0}). We obtain
a distance of \ve{7.97}{0.55} Mpc and \ve{7.49}{0.14} Mpc respectively. In case of free $t_0$, we arrived at
an explosion epoch of $-0.87$ days which is well within uncertainty and consistent with that known observationally.
Hence, in this case it is absolutely unnecessary to fix the explosion epoch.

\epm\ has been applied to this SN \citep{takats06} and a distance of \ve{7.1}{1.2} Mpc has been determined.
However for this a \ebv=0.11 has been used, the value of reddening was updated to 0.05 by \cite{pasto09} which
we adopt in our work, this accounts for the higher value of $ D $ estimated in this work.
\cite{2012A&A...540A..93V} has presented an improved distance estimate of \ve{8.4}{0.7} Mpc for the
host galaxy M51 by applying \epm\ on both 2005cs and 2011dh. Another \epm\ estimate for the SN has been
presented by \cite{dessart08} in which they derived distance of \ve{8.9}{0.5} Mpc. Both of
these \epm\ estimates are in good agreement with our results.

SN 2006bp:
Photometric data presented by \cite{quimby07} is available for $UBVri$ filter, but due to lack of $\xi$
prescriptions for SDSS filter, viz., \{BVi\} or \{Vi\}, $ri$ data may not be directly used. We have therefore,
carried out \epm\ analysis using \{BV\} subset only and the results are shown in Fig.~\ref{fig:epm_06bp}.
For D05 $\xi$ prescription, we derived a distance of $18.82\pm1.04$ Mpc.
\cite{dessart08} has also applied \epm\ on the SN in which they estimated the distance using re-computed set
of dilution factors and obtained a distance \ve{17.5}{0.8} Mpc which is consistent with our estimate within
the limit of errors.

SN 2008in:
We used photometric data at 21 epochs from \cite{roy11} to estimate \epm\ distance. The spectroscopic coverage
of the event is not very good specially within +50 day and hence we had to largely rely upon interpolation (see
Table~\ref{tab:vel-sne}). It is noteworthy to mention that we found the velocity of profile of the event is quite
less varying and overall velocities are much less as compared to other normal type events. This is also
supported by the fact that it is classified as spectroscopically sub-luminous \citep{roy11}.
The \epm\ fitting is shown in Fig.~\ref{fig:epm_08in} and we derive a distance of \ve{14.51}{1.38} Mpc
and $t_0$=\ve{-4.97}{1.91} days for D05 prescription. \cite{2013arXiv1306.5122U} has estimated the explosion
epoch for this event using hydrodynamical modelling and estimated an explosion epoch nearly 4 days prior
to our adopted reference epoch, thus this shows a very good agreement with EPM estimated $t_0$.

SN 2009bw:
Fig. \ref{fig:epm_09bw} shows the \epm\ fitting for this event. It is noted that even though we were having
photometric data starting from +5 day (see Fig.~\ref{fig:photplot_all}), but due to single spectra at +4d and
unavailability of any other
spectra before +18d, we could only include data points within +10 to +50 days for the \epm\ fit. This was
necessary to do as in early phase the velocity profile is quite steeper as compared to later phases and thus
in such phases velocity interpolation might go wrong due to less number of spectroscopic data.

Using both dilution factor prescription, we find that the distances derived using band sets \{BV\} and \{BVI\}
are very much consistent with each other, whereas the distance derived using \{VI\} subset is significantly
higher and the explosion epoch is also very much off (see Table \ref{tab:dis-epm}). Thus
making the \epm\ fit of \{VI\} quite inconsistent with the rest of two band-sets
and also the explosion epoch is not consistent with SN age estimated from spectra and light-curve evolution.
This particular inconsistency can be justified by the fact that \{VI\} band-set are at the cooler ends of SED
as compared to \{BV\} and \{BVI\} band-sets. As in early phase SED is hotter and estimation of SED parameters
viz., $ \theta $ and temperature, using \{VI\} band-set will be more prone to errors if the photometric
magnitude uncertainty is significant as we have in literature data of SN 2009bw.

Using D05 prescription, we derive a distance of $15.93\pm0.32$ Mpc. \cite{2009AJ....138..323T} derive
a distance of 11.1 Mpc to the host galaxy using Tully-Fisher method. No other redshift-independent
distance estimate is available for this galaxy. We, however, note that \cite{inserra12} adopted a
distance of 20.2 Mpc based on the redshift of the galaxy.

SN 2009md:
Figure \ref{fig:epm_09md} shows the \epm\ fit for this event. Extremely large errors in $\theta/v_{ph}$ quantity
can be noted and it is attributed entirely due to large photometric errors, as errors in photometric magnitudes have
amplified exponentially in fluxes and propagated all throughout to $\theta/v_{ph}$ quantities. Using D05
prescription we obtained a distance of \ve{23.29}{1.96} Mpc and the time of explosion of \ve{4.15}{1.97} days.
\cite{fraser11} has applied \scm\ to this event and derived a distance of 18.9 Mpc using optical data and
of 21.2 Mpc using near infrared data. For this case, the \epm\ result is consistent with that
derived using \scm.

SN 2012aw:
This is a well studied nearby event. The explosion epoch is known fairly accurate with an
error of $\pm0.79$ days, see \cite{bose13aw} and references therein. Figure \ref{fig:epm_12aw_fixt0}
and Fig.~\ref{fig:epm_12aw} show the \epm\ fit respectively with fixed and free $ t_0 $.
For D05 prescription, we derive a distance of \ve{9.83}{0.41} Mpc  and $11.27\pm0.88$ Mpc with explosion
$t_0=$\ve{-2.36}{0.75}. No previous \epm\, study exist for the galaxy NGC 3351, but recent
Cepheids \citep{2001ApJ...553...47F} and Tully-Fisher \citep{2002ApJ...565..681R} distance estimates
are in good agreement with our result.

\section{Summary} \label{sec:sum}
In this study we present \epm\ distances for eight host galaxies derived using photometric and
spectroscopic data of IIP SNe
The SNe have mid-plateau absolute V-magnitudes in the range -17 to -15. Detailed \epm\ analysis is done
for five of the events, viz., SN 2004et, 2008in, 2009bw, 2009md and 2012aw for the first time.
We use two dilution factor models, three filter sub-sets, and two methods for photospheric
velocity determination. The value of reddening are known quite accurately and for few of the events
the explosion epochs are constrained observationally with an accuracy of a day.
We find that \epm-derived distances using above two models differs
by 30-50\%. The \epm\ distances derived using Hamuy's model \citep{h01} are found to be systematically
lower than that of Dessart ones \citep{dh05}. For all the events in our sample, the distances using
Dessart model is found to be consistent with that derived using other redshift independent methods, i.e.
Tully Fisher, Standard Candle Method, Cepheid, Surface brightness fluctuation. We
also note that \epm\ method is applicable only to the early ($<$ 50 d) photometric data
of supernovae.

We have also studied the effect of two methods of velocity estimation on the derived distance. It is found that
the \synow\ model velocities are significantly different than that estimated by locating absorption trough
of P-Cygni. The distances derived from two different velocity determination methods have notable
differences as high as 15-18\%, however we did not find any systematic trend of this difference.
This suggests the difference is the direct effect of the measurement error of absorption minima method
when the photospheric lines are blended or weak relative to continuum.

\acknowledgments
We thank M. Fraser, C. Inserra, A. Pastorello and K. Takats for providing their spectra of respective SN
2009md, SN 2009bw, SN 2005cs and early spectra of SN 2004et. Their invaluable contribution has helped
immensely in preparation of this work and enriching our sample of SNe. We gratefully acknowledge the services
of the
NASA ADS and NED databases and also the online supernova spectrum archive (SUSPECT) which are used to access
data and references in this paper. Authors are also thankful to the referee whose thoughtful comments and
suggestions has significantly improved this work.

\bibliographystyle{apj}
\bibliography{ms}

\end{document}